\documentclass[11pt,a4paper]{article}
\usepackage[dvips]{graphicx}

\usepackage{epsfig}
\usepackage{epstopdf}
\usepackage{graphicx}
\usepackage{indentfirst}
\usepackage{amsmath}
\usepackage{amsfonts}
\usepackage{amssymb}
 \usepackage{textcomp}

\usepackage[pdftex,pdfmenubar=true,bookmarks=true,pdftoolbar=true]{hyperref}
\hypersetup{colorlinks,citecolor=blue,filecolor=magenta,linkcolor=magenta,urlcolor=cyan,pdftex}

\newcommand{\be}{\begin{equation}}
\newcommand{\ee}{\end{equation}}
\newcommand{\bea}{\begin{eqnarray}}
\newcommand{\eea}{\end{eqnarray}}

\begin{document}

\begin{center}
\begin{flushright}\begin{small}    
\end{small} \end{flushright} \vspace{1.5cm}
\Large{\bf Perfect fluid and  $F(T)$ gravity descriptions of inflationary universe and comparison with observational data
} 
\end{center}

\begin{center}
M. G. Ganiou $^{(a)}$\footnote{e-mail:moussiliou\_ganiou @yahoo.fr},
M. J. S. Houndjo   $^{(a,b)}$\footnote{ e-mail: sthoundjo@yahoo.fr},
I. G. Salako $^{(a,c)}$\footnote{ e-mail: inessalako@gmail.com},
 M. E. Rodrigues   $^{(d)}$\footnote{e-mail: esialg@gmail.com},
and J. Tossa $^{(a)}$\footnote{e-mail: joel.tassa@imsp-uac.org}
\vskip 4mm
$^a$ \,{\it Institut de Math\'{e}matiques et de Sciences Physiques (IMSP)}\\
 {\it 01 BP 613,  Porto-Novo, B\'{e}nin}\\
  $^{b}$\,{\it Facult\'e des Sciences et Techniques de Natitingou - Universit\'e de Parakou - B\'enin} \\
  $^{c}$\,{\it D\'epartement de Physique ,  Universit\'e d'Agriculture
de K\'etou, BP 13 K\'etou,  B\'enin}\\
$^{d}$\,{\it Faculdade de Ci\^encias Exatas e Tecnologia, Universidade Federal do Par\'a - Campus Universit\'ario de
 Abaetetuba, CEP 68440-000, Abaetetuba, Par\'a, Brazil}\\
\vskip 2mm
\end{center}

\begin{abstract}
\hspace{0,2cm} 
We describe in this paper the observables of inflationary models, in particular the spectrum index of torsion scalar perturbations, 
the tensor-to-scalar ratio, and the running of the spectral index, 
in the framework of perfect fluid models and  $F(T)$ gravity theories through the reconstruction methods. 
Then, our results on the perfect fluid and  $F(T)$ gravity theories of inflation are compared  with recent cosmological observations
 such as the Planck satellite and BICEP2 experiment. Ours studies prove that the perfect fluid and  $F(T)$ gravity models can reproduce the 
 inflationary universe consistent above all with the Planck data. We have reconstructed several models and considered others which 
 give the best fit values compatible with the spectral index of curvature perturbations, the tensor-to-scalar ratio,
 and the running of the spectral index within the allowed ranges suggested by the Planck and BICEP2 results.
\end{abstract}
\tableofcontents

\section{Introduction}

The recent data taken by the BICEP2 experiment ( Ade et al. 2014) on the tensor-to-scalar ratio of the primordial density
perturbations, additionally to the observations by the satellites of the Wilkinson Microwave Anisotropy Probe (WMAP)
( Spergel et al. 2003;  Hinshaw et al. 2013)
and the Planck(  Ade et al. 2014) have begotten many reflexion  on inflation. The potential form of inflaton is 
related to the spectrum 
of the density perturbations generated during inflation ( Lidsey et al. 1997). Several models of inflation have recently  
been constructed such as quantum cosmological perturbations for predictions and observations ( Mukhanov  2013), and others
 to account for the Planck and BICEP2 experiment ( Hazra et al. 2014). It exists some of them which are related to scalar field additionally
 to modified gravities 
 models of inflation in comparison with the data analysis of the BICEP2 ( Joergensen et al. 2014; Gao and Gong 2014;  Bamba et 
 al. 2014).\par
 Many studies with Various interesting results, have been done  on the reconstruction of inflationary models in 
 the framework of perfect fluid , $F(R)$ gravity and others 
 modified gravity theories. Furthermore,  the reconstruction 
of F (R) gravity models from observational data has been executed in  ( Starobinsky 1980). It has been also done in  supergravity ( Ferrara et al. 2014) and the related models 
( Chakravarty and Mohanty 2015). All these realizations  have been the attempts 
 to make modified gravity models to explain the Planck and BICEP2 results ( Bamba et al. 2014; Pallis 2014 ).  Moreover,  Bamba and his 
 collaborators have recently and explicitly performed the reconstruction of scalar field theories with  inflation leading
 to the theoretical consequences compatible with the observational data obtained from the Planck and BICEP2 in terms of 
 the spectral index of the curvature fluctuations, the tensor-to-scalar ratio, and the running of the spectral index in 
 (  Bamba et al. 2014). They have also made the perfect fluid and $F(R)$ gravity descriptions of inflation and its comparison with observational 
 data in ( Bamba et al. 2014). In this last one , they have re-expressed the observables of inflationary models, i.e., the spectral
 index $n_s$  of curvature perturbations, the tensor-to-scalar ratio $r$, and the running of the spectral index $\alpha_s$ ,
in terms of the quantities in perfect fluid models and $F (R)$ gravity theories. They have investigated several  $F (R)$ 
gravity models, especially, a power-law model which gives the best fit values compatible with the spectral index 
and tensor-to-scalar ratio within the allowed ranges suggested by the Planck and BICEP2 results. Moreover, the spectral index's 
features have been studied in induced gravity ( Kaiser 1994) and scalar-tensor theories ( Kaiser 1995). {\bf An important work which explains a 
scenario of inflation  for the first time in the framework of teleparallel gravity preciously  in  the modified  version
$F (T)$ gravity is proposed in  
(Jamil et Al.). Their investigation gived a value of Spectral scalar index $n_s$ which is compatible in a reasonable 
agreement with the WMAP7 and for pre Planck $ns < 0.961$}.\par 
 In parallel way  to these works with interesting results on the inflationary models in the framework of perfect fluid 
 and modified gravities,  we investigated in this paper, the descriptions of inflation through perfect fluid and
 $F(T)$ gravity models. We reformulate the observables  of inflationary models in terms of the quantities of scalar
 torsion and in perfect fluid  and $F (T)$ gravity models  and compare the theoretical representations 
 with recent data from  Planck et BICEP2. We here emphasize that  the $F (T)$ gravity coupling with scalar
 field theory have already been explored in several interesting cosmological works the  density perturbation
 growth in teleparallel cosmology ( Geng and Wu 2013) and  dynamical features of scalar-torsion theories ( Skugoreva et al. 20015). 
 This present work wants to explain the importance of formulations of the observables  for inflationary 
 models in terms of the perfect fluid and $F (T)$ gravity. It is physically motivated by the fact that 
 in ordinary scalar field model of inflation, the  spectral index, the tensor-to-scalar ratio, and the running of the spectral 
 index are represented by using the potential $V(\phi)$ of the scalar field. Consequently,  scalar field models are consistent with the
observations and by comparing the theoretical representations of theses observables with observations, 
we can get information on the properties of the perfect fluid and $F (T)$ gravity models to account
for the observations in the early universe. Our  approach in terms of inflation can permit to find the conditions in which 
the perfect fluid and $F (T)$ gravity models can be viable from the cosmological point of view. We normalize to unit the following constants 
$k_B=c=\hbar=1$ and express the gravitational constant  $8\pi G_N$ by $\kappa^2\equiv 8\pi /M^2_{PI}$ with Planck mass of 
$M^2_{PI}=G_N^{-1/2}$.\par
The plan of the manuscript is outlined as follows: In Sec.2 we write the slow-roll  parameters and express the observables for inflationary
models in the first time  as function of scalar torsion and in the second time in  perfect fluid description. Then, in Sec.3 we 
re-express these parameters and observables in framework of $F (T)$ gravity. All ours investigations have been ended by comparison with the 
Planck and BICEP2 data. Finally, the Sec.4 is devoted to the conclusions.  We present the explicit expressions of 
the slow-roll  parameters the observables for inflationary
models  as function of scalar torsion and  in the formulation of perfect fluid in Appendixes $A$ and $B$, respectively. The theoretical 
representations of observables for inflationary models in $F (T)$ gravity are presented in Appendix $C$. Appendix $D$ is devoted to 
the representations of the observables for inflationary models in the linear form  of the scalar torsion
and  in the exponential form. We have ended this work on Appendix $E$ where we have reminded the relation between the equation 
of state $(EoS)$  parameter and the tensor-to-scalar ratio. 

\section{Scalar torsion and perfect fluid description of the Slow-roll parameters}
As we have mentioned it in our introduction, the slow-roll parameters are related to inflaton  namely the potential of scalar field. In this 
section, we express these parameters by the scalar torsion and in terms of perfect fluid.
\subsection{ SLow-roll parameters}
 The action of teleparallel gravity coupled with the model of scalar field $\phi$ is given by ( Geng and Wu 2013  )
 \begin{equation}\label{s1}
   S=\int \left(\frac{T}{2\kappa^2}-\frac{1}{2}\partial_\mu \phi \partial^\mu \phi-V(\phi)\right)ed^4x,
 \end{equation}
where $T$  is the scalar torsion and $e$ the determinant of  tetrad  $e^{a}\;_\mu$ .
The slow-roll parameters, $\varepsilon$, $\eta$  et  $\xi$  are defined by (Bamba et al. 2014) 
\begin{equation}\label{sl}
 \varepsilon \equiv\frac{1}{2\kappa^2}\left(\frac{V^\prime(\phi)}{V(\phi)}\right)^2\quad ,\quad 
 \eta\equiv\frac{1}{\kappa^2}\frac{V^{\prime\prime}(\phi)}{V(\phi)}\quad ,\quad 
 \xi^2\equiv\frac{1}{\kappa^4}\frac{V^\prime(\phi)V^{\prime\prime\prime}(\phi)}{(V(\phi))^2}.
\end{equation}
Here and for the rest, the prime  means the derivative with respect to argument such as  
$V^\prime(\phi)\equiv \partial V(\phi)/\partial \phi$,  and others.
For the scalar field models, the spectral index $n_s$, of curvature perturbations, the tensor-to-scalar ratio $r$  of the density
perturbations and the running of spectral index  $\alpha_s$ are expressed as (Bamba et al. 2014) 
\begin{equation}\label{ob}
 n_s-1\sim -6\varepsilon+2\eta,\quad r=16\varepsilon, \quad \alpha_s\equiv \frac{dn_s}{dln\kappa}\sim16\varepsilon\eta-24\varepsilon^2-2\xi^2.
\end{equation}
 The variation of action (\ref{s1}) with respect to the tetrad   $e^{a}\;_\mu$ gives (  Geng and Wu 2013; Skugoreva et al. 2015) 
\begin{eqnarray}\label{s2}
 \frac{1}{\kappa^2}\left[  \frac{1}{4}\delta^{\rho}_{\beta}T +T^{\sigma}\;_{\nu\beta}S_{\sigma}\;^{\rho\nu} +
 e^{-1}e^{a}\;_{\beta}\partial_{\alpha}\left(ee_{a}\;^{\sigma}S_{\sigma}\;^{\rho\alpha}\right) \right]=\nonumber\\
 \frac{1}{2}\left[\delta^{\rho}_{\beta}(\frac{1}{2}g^{\mu\nu}\partial_\mu \phi \partial_\nu \phi+V(\phi)) -\frac{1}{2}\delta^{\mu}_{\beta}
  g^{\mu\rho}\partial_\mu \phi \partial_\nu \phi-\frac{1}{2}\delta^{\nu}_{\beta}
  g^{\nu\rho}\partial_\mu \phi \partial_\nu \phi \right].
\end{eqnarray}
We consider an universe described by the following flat Friedmann-Lema\^itre-Roberson-Walker FLRW metric:
\begin{equation}\label{s3}
 ds^2=-dt^2+a^2(t)\sum_{i=1,2,3} (dx^i)^2.
\end{equation}
Here, $a(t)$ is the scale factor and $H\equiv \dot{a}/a$ is the Hubble parameter. From (\ref{s3}), one obtains  the torsion scalar in function
of  $H$ by $T=-6H^2$. The point $(.)$ means here the derivative with respect to time as  $\partial/\partial t$.\\ 
 We can now find out the expressions of the slow-roll parameters as function of torsion scalar. Indeed, by using the relations (\ref{s2}) and
 (\ref{s3}), one gets the field equation given by: 
\begin{equation}\label{s4}
 \frac{T}{2\kappa^2}=-\frac{1}{2}\dot{\phi}^2 -V(\phi),
\end{equation}
\begin{equation}\label{s5} 
 \frac{\dot{T}}{12H\kappa^2}=\frac{1}{2}\dot{\phi}^2.
\end{equation}
 We define the scalar field $\phi$ by a new scalar field $\varphi$ such as $\phi=\phi(\varphi)$  and we identify  $\varphi$ to e.folds 
 number $N$. If one defines $N$ by scale factor as $N\equiv ln(a/a_0)$, we obtain $\dot{N}=\dot{\varphi}\equiv H(N)$. The equations 
  (\ref{s4}) and  (\ref{s3}) become  
\begin{equation}\label{s6}
 \frac{T(N)}{2\kappa^2}=\frac{1}{12}w(\varphi)T(N)+V(\phi(\varphi)),
\end{equation}

 \begin{equation}\label{s7}
  \frac{T^\prime(N)}{12\kappa^2}=-\frac{1}{12}w(\varphi)T(N),
 \end{equation}
with $w(\varphi)\equiv(d\phi/d\varphi)^2$. So, if we consider  the torsion scalar as function  of $N$, by combining (\ref{s6}) and  (\ref{s7}),
 one expresses  $w(\varphi)$ and $V(\phi)\equiv V(\phi(\varphi))$ by $N$ as   

\begin{eqnarray}\label{s8}
 w(\varphi)=[\frac{-T^\prime(N)}{\kappa^2T(N)}]_{N=\varphi} , 
\end{eqnarray}
\begin{equation}\label{s9}
 V(\varphi)= -\frac{1}{12\kappa^2}\left[6T(N)+  T^\prime(N)\right]_{N=\varphi}.
\end{equation}
 We  find $T=T(N)$ and  $\varphi=N$ as solution for field equation   $\phi$ or   $\varphi$ and  Einstein equations because of equivalence 
 between Teleparallel and General Relativity. We also find that $T^\prime(N)>0$ because  $w(\varphi)>0$ and $T(N)<0$. {\bf Thus, we can now 
 express in terms of $T(N)$ all the slow-roll parameters and the observables  by making using the relations (\ref{s8})
 and (\ref{s9}) in their basic expressions defined in  (\ref{sl}) and (\ref{ob}) respectively. One gets  
 \begin{eqnarray}
 \epsilon&=& -\frac{T(N)}{2T^\prime(N)}\left(\frac{6T^\prime(N)+T^{\prime\prime}(N)}{6T(N)+T^\prime(N)}\right)^2\\
 \cr \eta&=& -[6(T'(N))^3+T^{\prime\prime}(N)(T'(N))^2+6T^{\prime\prime}(N)T'(N)T(N)-(T^{\prime\prime}(N))^2T(N)\\
&&+2T^{\prime\prime\prime}(N)T'(N)T(N)]/[2(6(T'(N))^2T(N)+(T'(N))^3)]
 \cr \xi^2&=& \frac{T(N)}{2T^\prime(N)}\frac{6T^\prime(N)+T^{\prime\prime}(N)}{(6T(N)+T^\prime(N))^2}
[7T^{\prime\prime}(N)+3T^{\prime\prime\prime}(N)+\frac{2T^{\prime\prime\prime\prime}(N)T(N)}{T^\prime(N)}+
\frac{6T^{\prime\prime\prime}(N)T(N)}{T^\prime(N)}\nonumber\\
 &&-\frac{4T^{\prime\prime\prime}(N) T^{\prime\prime}(N)(N)}{(T^\prime(N))^2}+\frac{2 (T^{\prime\prime}(N))^3T(N)}{(T^\prime(N))^3}-
\frac{6(T^{\prime\prime}(N))^2T(N)}{(T^\prime(N))^2}-
\frac{ (T^{\prime\prime}(N))^2}{T^\prime(N)}]
\end{eqnarray}
\begin{eqnarray}
\label{salai}
n_s&=&1+\frac{3 T(N) \left(6 T'(N)+T''(N)\right)^2}{T'(N) \left(6 T(N)+T'(N)\right)^2}\nonumber\\
&&+\frac{-6 T'(N)^3-T'(N)^2 T''(N)+T(N) T''(N)^2-2 T(N)
T'(N) (3 T''(N)+T^{(3)}(N))}{T'(N)^2 (6 T(N)+T'(N))},
\end{eqnarray}
\begin{eqnarray}
\label{sort}
 \cr r&=&-\frac{8 T(N) \left(6 T'(N)+T''(N)\right)^2}{T'(N) \left(6 T(N)+T'(N)\right)^2},
 \end{eqnarray}
\begin{eqnarray}
\label{alf}
\cr \alpha_s&=& [T(N) (6 T'(N)+T''(N)) (T'(N)^4 (144 T'(N)^2+5 T''(N)^2+T'(N) (41 T''(N)
-3 T^{(3)}(N)))\nonumber\\
 &&+12T(N)^2 T'(N) (-4 T''(N)^3+T'(N) T''(N) (9 T''(N)+8 T^{(3)}(N))+T'(N)^2 (51 T''(N)+9 T^{(3)}(N)\nonumber\\
&&-2 T^{(4)}(N)))-2T(N) T'(N)^2 (216 T'(N)^3+6 T''(N)^3+3 T'(N) T''(N) (11 T''(N) -2 T^{(3)}(N))\nonumber\\
&&+ T'(N)^2 (150 T''(N) -3 T^{(3)}(N)+T^{(4)}(N)))-72T(N)^3 (T''(N)^3-T'(N) T''(N) (3 T''(N)\nonumber\\
&&+2 T^{(3)}(N))+T'(N)^2 (3 T^{(3)}(N)+T^{(4)}(N))))]/(T'(N)^4(6 T(N)+T'(N))^4).
\end{eqnarray} }
 \subsection{Perfect fluid  description of slow-roll parameters and observables }
  We rewrite here, the slow-roll parameters in framework of perfect fluid. According to the FLRW metric (\ref{s3}), the gravitational field 
  equations for a perfect fluid take the following forms ( Geng and Wu 2013 )
  \begin{equation}\label{s10}
 \frac{T(N)}{2\kappa^2}=-\rho(N),
\end{equation}
 \begin{equation}\label{s11}
  \frac{T^\prime(N)}{6\kappa^2}=\rho(N)+P(N),
 \end{equation}
where $\rho(N)$ and  $P(N)$ are energy density and pressure of perfect fluid respectively. Suppose the following  general equation of state:
  \begin{equation}\label{s12}
   P(N)=-\rho(N) +f(\rho),
  \end{equation}
with $f(\rho)$ a function of $\rho$. In this case, the equation (\ref{s11}) takes the form 
  \begin{equation}\label{s13}
  \frac{T^\prime(N)}{6\kappa^2}=f(\rho).
 \end{equation}
 The conservation law  $\rho^\prime(N)+3[\rho(N)+ P(N)]=0$ also  becomes 
\begin{equation}\label{s14}
   \rho^\prime(N)+3f(\rho)=0.
  \end{equation}
  By combining equations  (\ref{s13}) and  (\ref{s14}), we obtain 
\begin{equation}\label{s15}
    \frac{T^{\prime\prime}(N)}{6\kappa^2}=-3\tilde{f}(\rho)f(\rho).
  \end{equation}
 We emphasize here that the tilde of  $\tilde{f}$ means its derivative with respect to $\rho$ i.e $\tilde{f}(\rho)\equiv df(\rho)/d\rho$, 
 whereas the prime means the derivative with respect to  $N$ i.e   $T^\prime(N)\equiv dT(N)/dN$ and $\rho^\prime(N)\equiv d\rho(N)/dN$.
 {\bf By using the relations  (\ref{s8}), (\ref{s9}), (\ref{s10}) and  (\ref{s13}), we can re-write the slow-roll parameters
 and the observables as only 
 function of  $\rho(N)$ and $f(\rho(N))$ as it was done in ( Bamba et Al.  2014).}
\subsection{Perfect fluid models reconstructions}
 We study  two forms of scalar torsion to reconstruct two models of perfect fluid: the linear form and the exponential form.

 \subsubsection{ The linear form }
 We investigate here the linear form of  $T(N)$ given  by:    
\begin{equation}\label{s16}
 T(N)=D_0N+ D_1,
\end{equation}
with  $D_0>0$ and  $D_1<0$ constants. The physical motivation of this choice is that, according to exponential inflation of slow-roll, the 
scale factor is given by $a=\bar{a}\exp({H_{int}t})$ where $\bar{a}$ is a constant (Bamba et al. 2014).  
$H_{int}$ is the Hubble parameter at
the inflationary stage and it is approximately constant i.e it weakly depends of time. To express this weak time dependence of $H$ and 
consequently of scalar torsion $T$ we use the form in (\ref{s16})  where the e-fold number plays the role of time. In this case,  
if  $D_1/D_0\ll N$, the 
time dependence of scalar torsion during inflation is negligible. By using the  relations (\ref{s10}),  (\ref{s11}), one obtains: 
\begin{equation}\label{s17}
  \rho(N)=-\frac{1}{2\kappa^2}(D_0N+ D_1), \quad P(N)=\frac{1}{6\kappa^2}[(3N+1)D_0+3D_1].
 \end{equation}   
By eliminating $N$ between these last equations, we  find the following relation 

\begin{equation}
  P(N)=\frac{D_0}{6\kappa^2}-\rho(N),
\end{equation}
which, added to the general equation of state (\ref{s12}), gives 

\begin{equation}
 f(\rho)= \frac{D_0}{6\kappa^2}.
\end{equation}
From equations in (\ref{s17}), we can deduce the state equation of perfect fluid according to the linear form of scalar torsion as 

\begin{equation}\label{s18}
 \omega(N)=\frac{P(N)}{\rho(N)}=-1+\frac{f(\rho)}{\rho(N)}=\frac{(3N+1)D_0+3D_1}{3(D_0N+D_1)}.
\end{equation}

\subsubsection{The exponential form}
The second example of perfect fluid model studied in this work is whose scalar torsion is given by the following exponential
function of $N$

\begin{equation}\label{s19}
 T(N)=D_2Ne^{\beta N}+D_3,
\end{equation}

with  $D_2>0$, $D_3<0$ and $\beta >0$  constants.
The physical reason of the choice of this form is the following. During power law inflation, the scale  factor is given by 
$a=\bar{a}t^{\hat{p}}$ with $\hat{p}$  constant.  Then, the scalar torsion during inflation is so  $T=-6(\hat{p}/t)^2$ and becomes 
 $T=-6\hat{p}^2\exp(-2N/\hat{p})$. This last form is equivalent to (\ref{s19}) if we make $D_2=-6\hat{p}^2$, $\beta=- 2/\hat{p}$
 and $D_3=0$. Such an exponential form can reproduce the power-low inflation.\\
 By introducing relation (\ref{s19}) in the gravitational equations (\ref{s10}) and  (\ref{s11}), one gets

\begin{equation}\label{s19p}
 \rho(N)=-\frac{1}{2\kappa^2}(D_2Ne^{\beta N}+D_3), \quad   P(N)=\frac{1}{6\kappa^2}[(3+\beta)D_2Ne^{\beta N}+3D_3].
\end{equation}
 Elimination of $N$  between these equations gives  
\begin{equation}
  P(N)=-(1+\frac{\beta}{3})\rho(N)-\frac{\beta D_3}{6\kappa^2}.
\end{equation}
From equation (\ref{s12}), one obtains: 
\begin{equation}
 f(\rho)= \frac{\beta \rho(N)}{3} -\frac{\beta D_3}{6\kappa^2}.
\end{equation}
 Then, we find the parameter of state of perfect fluid model corresponding to the exponential form

\begin{equation}\label{s20} 
 \omega(N)=\frac{(3+\beta)D_2Ne^{\beta N}+3D_3}{3(D_2Ne^{\beta N}+D_3)}.
\end{equation}
\subsubsection{Another form}
We consider here a model studied in (Mukhanov 2013)  whose state parameter is given by 

\begin{equation}\label{s21}
 \omega(N)=-1+\frac{\bar{\beta}}{(1+N)^{\bar{\gamma}}},
\end{equation}

with $\bar{\beta}$ and $\bar{\gamma}$ free parameters. By solving the system of equations formed by (\ref{s10}), (\ref{s11}) et (\ref{s21}),
 we obtain 

\begin{equation}\label{bou}
 T=\bar{T}\exp\left[  \frac{-3\bar{\beta}(1+N)^{1-\bar{\gamma}}}{1-\bar{\gamma}}                  \right]
\end{equation}
where  $\bar{T}$ is constant.
{\bf
We  can now specify the  expressions of some observables as example the spectral index and the tensor-to-scalar ratio by introducing 
(\ref{bou}) in (\ref{salai}) and (\ref{sort}) }
\begin{eqnarray}
 n_s=\frac{1}{3 \left(-2 (1+N)^{\gamma }+\beta \right)^2}(1+N)^{-2-\gamma } [-9 (1+N)^2 \beta ^3+3 \beta ^2 (1+N)^{1+\gamma }  (13+13 N-\gamma
)\nonumber\\-2 (1+N)^{2 \gamma } \beta  (24 (1+N)^2-\gamma +\gamma ^2)
+2 (1+N)^{3 \gamma }(6+6 N^2-(-4+\gamma ) \gamma +6 N (2+\gamma ))],
\end{eqnarray}

\begin{eqnarray}
r=(\frac{8 (1+N)^{-2-\gamma } \beta  \left(3 (1+N) \beta +(1+N)^{\gamma } (-6-6 N+\gamma )\right)^2}{3 \left(-2 (1+N)^{\gamma }+\beta 
\right)^2}.
\end{eqnarray}
We deduce that for the appropriate value of parameters  $\{\bar{\beta},\bar{\gamma}\}$, we can  get  the values of observables $n_s$ and $r$.

 \subsubsection{Comparison with the observations}
Suppose that the time variation of  $f(\rho)$ et $\rho$ during inflation is sufficiently small and that inflation is almost exponential as 
$\omega(N)\equiv P(N)/\rho(N)=-1+f(\rho)/\rho(N)\approx-1$,i.e.,$|f(\rho)/\rho(N)|\ll1$, {\bf one gets from expressions obtained 
 in subsection $2.2$ after writting the slow-roll parameters and the observables in  term of perfect fluid, 
 the following approximatif  relations.}

 \begin{equation}\label{apt}
  n_s\sim1-6\frac{f(\rho)}{\rho(N)},\quad r\approx24\frac{f(\rho)}{\rho(N)},\quad \alpha_s\approx-9\left(\frac{f(\rho)}{\rho(N)}\right)^2.
 \end{equation}
 We present here the recent observations on spectral index  $n_s$, the tensor-to-scalar ratio  $r$ and the running of spectral index
 $\alpha_s$. The recent data of Planck satellite  (  Ade et al 2014) suggested  $n_s=0.9603\pm0.0073(68\% CL)$,$r<0.11(95\% CL)$, and 
 $\alpha_s=-0.0134\pm0.0090(68\% CL)$[Planck et WMAP ( Spergel et al. 2003;  Hinshaw et al. 2013)]], whose  negative sign is at $1.5\sigma$. 
 The BICEP2 experiment (Ade et al. 2014) 
 implies $r=0.20^{+0.07}_{-0.05}(68\% CL)$. It is mentioned that discussions exist on how to subtract  the foreground, for example in  
 ( Ade et al 2015;  Kamionskowski and Kovetz 2014 ). Recently, progress appear also in  ( Colley and Gott 2015) 
 to ensure  the BICEP2 declarations.
 It has been also remarked that the representation 
 of $\alpha_s$ is also given in ( Bassett et al. 2006).\par
 From equation (\ref{apt}), we can see that when the condition   $f(\rho)/\rho(N)=6.65\texttimes 10^{-3}$
 is realized at the inflationary era,
 we find $(n_s,r,\alpha_s)=(0.960,0.160,-3.98\texttimes 10^{-4})$. In the  case of linear form of $T$  from equation  (\ref{s16}), if 
$D_1/D_0\gg N$ and $-D_0/(3D_1)=6.65\texttimes 10^{-3}$, the change of value of   $\omega(N)$ is consider to be negligible, and the above 
condition can be met at the inflationary stage. In the second time, for the exponential form with the condition  
$\beta=2.0\texttimes 10^{-4}$ 
 and then  $\beta N\ll1$, and that $-1/3\{[ 1+D_3/(D_2,_\beta)]\}=6.65\texttimes 10^{-3}$, $\omega(N)$ can be seen as a constant and above
  condition can be satisfied during inflation. As consequence, the conclusion is that perfect fluid can lead to Planck results  with 
  $r=\mathcal{O}(0.1)$, which value is compatible with  BICEP2 experiment. \\
  Thus, we mention here some concrete perfect fluid models  ( Astashenok et al. 2012) . For model of   $P(N)=-\rho(N)+f(\rho)$ with 
  $f(\rho)=\bar{f}\sin(\rho(N)/\bar{\rho})$ where  $\bar{f}$ is constant and $\bar{\rho}$, the fiducial value of $\rho$ known to 
  produce the scenario of Pseudo-Rip ( Frampon et al. 2011), if $\rho((N)/\bar{\rho}\ll 1$ and 
  $f(\rho)/\rho(N)\approx \bar{f}/\bar{\rho}=6.65\texttimes 10^{-3}$ behaves almost  constant and the above condition can be satisfied. Then,
  We have  examined another model of $P(N)=-\rho(N)+f(\rho)$, where $f(\rho)=(\rho(n))^\tau$  with $\tau(\neq0)$ constant. By using equation
  (\ref{s10}), one gets $f(\rho)/\rho(N)\approx (-T_{inf}/\kappa^2)^{\tau-1} $, where the constant $T_{inf}$ means the scalar torsion at the 
  slow-roll inflation regime. So, similarly to the first example, when 
  $f(\rho)/\rho(N)\approx (-T_{inf}/\kappa^2)^{\tau-1} =6.65\texttimes 10^{-3}$, $f(\rho)/\rho(N)$ can be considered constant and the above   
  condition can be met.

  \section{Description of the SLOW-ROLL  in $F(T)$ gravity}
We describe in this section, the slow-roll parameters in the framework of $F(T)$ gravity. 

\subsection{ Observables of inflationary models in terms of the quantities in   $F(T)$ gravity theories }
The action in context of  $F(T)$ gravity theories is defined by
 \begin{eqnarray}\label{actio}
  S_{F(T)}=\int e\left(\frac{F(T)}{2\kappa^2} + \mathcal{L}_{matter}\right)dx^4, 
 \end{eqnarray}
The equation of motion is determinated by ( Salako et al 2013)
\begin{eqnarray}
S_{\beta}\;^{\rho\alpha}\partial_{\alpha}Tf_{TT}+\left[e^{-1}e^{a}\;_{\beta}\partial_{\alpha}\left(ee_{a}\;^{\sigma}S_{\sigma}\;^{\rho\alpha}\right)+
T^{\sigma}\;_{\nu\beta}S_{\sigma}\;^{\rho\nu}\right]f_{T}+\frac{1}{4}\delta^{\rho}_{\beta}f=4\pi\mathcal{T}^{\rho}_{\beta},
\end{eqnarray}
 where $\mathcal{T}^{\rho}_{\beta}$ represents energy momentum tensor of matter. By using the metric  $FLRW$ in  (\ref{s3}), the modified
 Friedmann equations take the forms 
 
 \begin{eqnarray}
\label{e1}
 \frac{1}{2}F(T)-TF^{\prime}(T)+\kappa^2\rho_{matter}=0, \label{e1}\\
 \label{e2}
 \frac{1}{2}F(T)+(6H^2+2\dot{H})F^{\prime}(T)-24H^2\dot{H}F^{\prime \prime}(T)-\kappa^2P_{matter}=0. 
\end{eqnarray}
$\rho_{matter}$ et $P_{matter}$ are respectively energy density and the pressure of matter. The prime of $F(T)$ means it derivative with 
respect to scalar torsion $T$ as example   $F^{\prime}(T)=\frac{dF(T)}{dT},\quad F^{\prime \prime}(T)=\frac{d^2F(T)}{dT^2} $, etc. 
  In vacuum, the equation  (\ref{e1}) gives

 \begin{eqnarray}\label{tf}
  T=\frac{F(T)}{2F^{\prime}(T)}.
 \end{eqnarray}
With the fact that   the scalar tension $T$ is  function of  $N$, by introducing the relation  (\ref{tf}) in the equations
(\ref{s10}) and (\ref{s13}), one gets $\rho(N)$ and  $f(\rho)$ in the contest of  $F(T)$ gravity. All  the slow-roll parameters  
and the observables  can be expressed as function of $F(T)$.
 
 \subsection{Reconstruction of  $F(T)$ gravity models }
  We use the method of reconstruction as it is developed in ( Nojiri and Odintsov 2006). 
  We define the e-folds number as $N\equiv-ln(a/a_0)$ with $a_0$
 the scale factor at present time $t_0$, and the redshift becomes $z\equiv a_0/a-1$. Consequently, we have  $N=-ln(1+z)$. We write 
 the scalar torsion in terms of $N$ through the function  $D(N)$ as 
  \begin{eqnarray}\label{s23}
   T(N)=D(N)=D(-ln(1+z))
  \end{eqnarray}
 By using  the relation  (\ref{s23}) and the continuity equation $\dot{\rho}_i+3H(1+\omega_i)\rho_i=0$, the equation (\ref{e2}) 
 can be  rewritten as
 \begin{eqnarray}\label{s24}
  0=&&-3F(T)+\left[6D(N)+D^{\prime}(N)\right]F^{\prime}(T)+2D(N)D^{\prime}(N)F^{\prime\prime}(T)\nonumber\\
  &&+6\sum_i\rho_{matter_{0i}}\omega_ia_0^{-3(1+\omega_i)}\exp[-3(1+\omega_i)N(T)],
  \end{eqnarray}
 where the last term is equal to the total pressure of matter $P_{matter}$. Here, the matters are supposed to be fluid  labeled by 
 $ \textless i\textgreater$ with the same equation of state $\omega_i\equiv P_{matter_i}/\rho_{matter_i}$ where  $\rho_{matter_i}$ 
 and $P_{matter_i}$ are energy density and the pressure of $i^{th}$ fluid and  $\rho_{matter_{0i}}$ a constant. It follows that 
for any form of scalar tensor i.e the function   $D(N)$, the differential  equation   (\ref{s24}) can be resolved and the corresponding
gravitational action $F(T)$ (the model) which reproduces the expansion history described by scalar torsion  via the Hubble parameter $H$. 
Moreover, for particular $F(T)$  model , the function  $D(N)$ can also be obtained.  Then, with the expressions of observables 
of inflationary models described in subsection  $2.1$, we can get the corresponding predictions on the inflation for this particular
model $F(T)$.

 \subsubsection{ The linear form}
 We use the linear form of scalar tensor $T$ of the relation (\ref{s16}). This prove that the e-folds number can also be expressed in terms of 
 the scalar torsion as $N(T)=(T-D_1)/D_0$. {\bf The Friedmann equation  (\ref{s24}), in the vacuum, becomes second order differential equation in 
 $T$ labeled by
 \begin{eqnarray}\label{diff1}
 -3F(T)+(6T+D_0)F^\prime(T)+2TD_0F^{\prime\prime}(T)=0,
 \end{eqnarray}
 whose resolution gives}
 \begin{eqnarray}\label{s25}
 F(T)= C_1\sqrt{-T} +C_2\sqrt{-T}  \left(-\frac{2 e^{\frac{3 T}{D_0}}}{\sqrt{-T}}-\frac{2 \sqrt{3 \pi } 
 \text{Erf}\left[\frac{\sqrt{3}
\sqrt{-T}}{\sqrt{D_0}}\right]}{\sqrt{D_0}}\right).
\end{eqnarray}
$C_1$ and $C_2$ constants of integration whereas  $\text{Erf}[z] $ is the Gauss integral distribution given by
$\text{erf}[z] =\frac{2}{\sqrt{ \pi } }\int_0^ze^{-t^2}dt$.{\bf 
We have consequently reconstructed the gravitational Lagrangian density which can genrate the required expansion for any given 
 expansion history $T$ or for any given  e-folds number $N$.  To putting out the  
observables of inflationary models corresponding to the expansion history form  considered, one puts the relation (\ref{s16}) 
in those defined  by (\ref{salai}), (\ref{sort}) and (\ref{alf}) and gets 
\begin{eqnarray}
\label{of}
  n_s&=& 1+\frac{6 D_0 [12 D_1+D_0 (-1+12 N)]}{(D_0+6 D_1+6 D_0 N)^2},
  \end{eqnarray}
\begin{eqnarray}
\label{off}
 \cr r&=&-\frac{288 D_0 (D_1+D_0 N)}{(D_0+6 D_1+6 D_0 N)^2},
 \end{eqnarray}
\begin{eqnarray}
\label{offf}
 \cr \alpha_s&=&-\frac{864 D_0^2 (D_1+D_0 N) [3 D_1+D_0 (-1+3 N)]}{(D_0+6 D_1+6D_0 N)^4}.
\end{eqnarray}}
 The inflationary phase must last enough to account for initial conditions problems as example the so called problem of horizon and flatness.
 The value of e-folds number at the end of inflation must be $N_e\gtrsim 50$. {\bf The slow-roll parameters  
 should be smaller than unity during 
 inflation; whereas  they became  larger than or equal to unity at the end of inflation, $N=N_e$. As examples, if  
 $(N,D_0,D_1)=(50.0,0.850,-95.0)$ and  $(60.0,0.950,-115)$,  one gets  from the previous relations (\ref{of}), (\ref{off}), and (\ref{offf})
  the following value of the observables $(n_s,r,\alpha_s)=(0.967,0.130,-5.32\times 10^{-4})$ and
 $(0.967,0.131,-5.45\times 10^{-4})$ respectively. Thus, the Planck results for  $n_s$ with $r=\mathcal{O} (0.1) $ can be reached. }

 \subsubsection{ The exponential form}
 Here, we use the exponential form of the scalar torsion in (\ref{s19}). In this case, the relation between   $N$ and the scalar torsion 
 is expressed  by $D_2e^{\beta N}= T-D_3$.{\bf Then, the second order differential equation obtained from   the Friedmann equation
 (\ref{s24}) in the vacuum according to this form of scalar torsion  is given by: 
 \begin{eqnarray}
  -3F(T)+(6T+\beta T-\beta D_3 )F^\prime(T)+2T(\beta T-\beta D_3)F^{\prime\prime}(T)=0.  
 \end{eqnarray}
Its resolution gives} 
\begin{eqnarray}\label{s26}
F(T)= C_1\sqrt{-T}-2C_2  (-D_3-T)^{-3/\beta } \left(1+\frac{T}{D_3}\right)^{3/\beta }\,   _2F_1\left[-\frac{1}{2},\frac{3}{\beta
},\frac{1}{2},\frac{-T}{D_3}\right],
\end{eqnarray}
 with $C_1$ and  $C_2$ the constants of integration and  $\text{2F1}(x_1,x_2,x_3; y ) $, is  the hypergeometric function where 
 $x_i (i=1...,3)$  are constants and $y$ the variable. 
 The corresponding observables of inflationary models for the exponential expansion history are given by 
 \begin{eqnarray}
 n_s&=&1+\frac{\beta  (6+\beta )\{-6 D_3^2+e^{N \beta } D_2 \left[2 D_3 \beta +e^{N \beta } D_2 (6+\beta )\right]\}}{\left[6
D_3+e^{N \beta } D_2 (6+\beta )\right]^2},\\
 \cr r&=& -\frac{8 e^{N \beta } D_2 \left(e^{N \beta } D_2+D_3\right) \beta  (6+\beta )^2}{\left(6 D_3+e^{N \beta } D_2
(6+\beta )\right)^2},\\ 
\cr \alpha_s&=&\Big \{e^{N \beta } D_2 (e^{N \beta } D_2+D_3) \beta ^2 (6+\beta ) \Big[5 e^{2 N \beta } D_2^2 (6+\beta)^2 
-2 e^{N \beta } D_2 D_3 (6+\beta )\nonumber\\
&&\times (-66+\beta ^2)+12 D_3^2 [51+2 \beta  (9+\beta )]\Big]\Big\}/[6 D_3+e^{N
\beta } D_2 (6+\beta )]^4.
\end{eqnarray}
{\bf We deduce from these  that for  $(N,D_2,D_3)=(50.0,1.10,-10)$ and $(60.0,1.20,-15.0)$, one gets 
 $(n_s,r,\alpha_s)=(0.9627,6.89\times 10^{-2},-6.4\times 10^{-4})$  and $(0.9652,5.83\times 10^{-2},-5.23\times 10^{-4})$ respectively.\par
 We recall here that our investigation tries to explain the inflation scenario in  the framework of $F(T)$ gravity coupled with scalar
 field as it was already done in a lot of works and also through another gravity ( Bamba et al. 2014 ; Jamil et al. 2015 ). The obtained 
 models in (\ref{s25}) and (\ref{s26}) represent the geometrical part of the initial model descibed by the action in (\ref{actio}).
 The reconstruction approach followed to obtain these models which consists to neglect all matter fields,  can also be justified by the 
  fact that in the inflationary era the inflation is driven by inflaton field $\phi$ (Jamil et al. 2015). Furthermore, for appropriate 
  choice of the parameters of these models, they can lead to the very interesting models studied in (Jamil et al. 2015).} 

\subsection{ Power-law model of $F(T)$ gravity }
In this part of  section, we  choose  one concrete model to describe the observables of inflationary models.the concerning model is 
the power-law model of  $F(T)$ gravity given by  $F(T)=\mu(-T)^n$ where $\mu$ are  $n$ constants. By introducing  this expression in the  Friedmann equation 
(\ref{s24}) in the vacuum, we get a first order differential  equation in $D(N)$ which solution is:
\begin{eqnarray}\label{lab}
   D(N)=T(N)= D_4\exp(-\frac{3N}{n}),
  \end{eqnarray}
with $D_4$ a negative constant. This  reconstructed form is equivalent to exponential form in (\ref{s19}) with the following conditions
$D_2=D_4$, $D_3=0$ and  $\beta=-3/n$. This solution is valid for  $n\neq0$. The slow-roll parameters with respect to (\ref{lab}) are 
written as $\epsilon=3/(2n)$, $\eta=3/n$, and $\xi^2=9/n^2$. Since these parameters are constants, if $n\gg1$, on gets 
$\epsilon\ll1$, $\eta\ll1$ and $\xi^2\ll1$ during inflation. These results prove the conditions of slow-rolling are realized 
in $F(T)$ gravity.


 \section{Conclusion}
 We have described in this paper the observables of inflationary models. These observables are related to the slow-roll parameters which 
 explicitly depend from the potential of inflaton. This justifies the choose of the scalar field in the description  of the observables. We 
 have find out the expressions of these observables as function of scalar torsion. This description have been done  in the framework of 
 scalar field model coupled with the Teleparallel gravity. We have chased ours investigations  on the perfect fluid description of slow-roll 
 parameters . After expressing the slow-roll parameters and the observables in terms of quantities of perfect fluid, we have reconstructed 
two perfect fluid models according to the linear and exponential forms of scalar tensor and gives additional model. The have compared the 
results with observation. We have found that perfect fluid can lead to the Planck results with $r=\mathcal{O}(o.1)$.\par
We have ended our study on the $F(T)$ gravity description of the slow-roll parameters. We  have also reconstructed  two models as we have  done 
in the case of perfect fluid. By choosing appropriate values for the free parameters of the models, one gets results on the observables
especially $n_s$ and $r$ which are compatible with the  Planck data on these observables. The last studied model in this work, the  
Power-law model of $F(T)$ gravity, has permitted to confirm the realization of the conditions of slow-rolling in the framework of  
$F(T)$ gravity.
 \vspace{1cm}
\newpage
 {\bf Acknowledgments}: 
 M.J.S.Houndjo thanks ``Ecole Normale Sup\'erieure de Natitingou" for partial financial support. M.E.Rodrigues also expresses
 his sincere gratitude to UFPA and CNPq.
\begin{center}
 \rule{8cm}{1pt}
\end{center} 

\end{document}